\RequirePackage{lineno} 
\documentclass[aps,twocolumn,secnumarabic,nobalancelastpage,amsmath,amssymb,nofootinbib,linenumbers]{revtex4}
\pdfoutput=1 

\usepackage{blindtext}
\usepackage{graphics}      
\usepackage{graphicx}      
\usepackage{longtable}     
\usepackage{url}           
\usepackage{bm}    
\usepackage{braket}        
\usepackage{xcolor}
\usepackage{float}         

\newcount\myloopcounter
\newcommand{\repeatit}[2][10]{
  \myloopcounter0 
  \loop\ifnum\myloopcounter < #1 
  #2
  \advance\myloopcounter by 1
  \repeat 
}

\begin{document}

\title{Observation of Domain Wall Confinement and Dynamics\\ in a Quantum Simulator}
\author{W. L. Tan\footnote{Correspondence to: wltan93@terpmail.umd.edu (W.L.T.); pbecker1@terpmail.umd.edu (P.B.)}, P. Becker${}^*$, F. Liu, G. Pagano, K. S. Collins, A. De, L. Feng, H. B. Kaplan, A. Kyprianidis, R. Lundgren, W. Morong, S. Whitsitt, A. V. Gorshkov, and C. Monroe}
\affiliation{Joint Quantum Institute and Joint Center for Quantum Information and Computer Science, University of Maryland and NIST, College Park, Maryland 20742, USA}
\date{\today}

\begin{abstract}
Confinement is a ubiquitous mechanism in nature, whereby particles feel an attractive force that increases without bound as they separate. A prominent example is color confinement in particle physics, in which baryons and mesons are produced by quark confinement. Analogously, confinement can also occur in low-energy quantum many-body systems when elementary excitations are confined into bound quasiparticles. Here, we report the first observation of magnetic domain wall confinement in interacting spin chains with a trapped-ion quantum simulator. By measuring how correlations spread, we show that confinement can dramatically suppress information propagation and thermalization in such many-body systems. We are able to quantitatively determine the excitation energy of domain wall bound states from non-equilibrium quench dynamics.  Furthermore, we study the number of domain wall excitations created for different quench parameters, in a regime that is difficult to model with classical computers. This work demonstrates the capability of quantum simulators for investigating exotic high-energy physics phenomena, such as quark collision and string breaking.

\end{abstract}
\maketitle

Fundamental constituents of matter, such as quarks, cannot be observed in isolation, because they are permanently confined into bound states of mesons or baryons. Although the existence of confinement in particle physics is well established, quantitative understanding of the connection between theoretical prediction and experimental observation remains an active area of research \cite{Greensite2011, Brambilla2014}. Similar phenomena can occur in low-energy quantum many-body systems, which can provide insight for understanding confinement from a microscopic perspective. The static and equilibrium properties of such confined systems have been well characterized in previous theoretical \cite{McCoy1978,Delfino1998,Fonseca2006}  as well as experimental works \cite{Lake2010,Coldea2010}. However, recent theoretical studies have demonstrated that confinement can also have dramatic consequences for the out-of-equilibrium dynamics of quantum many-body systems, such as suppression of information spreading and slow thermalization \cite{Kormos2017,Lerose2019,James2019,Fangli2019,Mazza2019,Lerose2019b,Verdel2019}.

\begin{figure}[th!]
    \includegraphics[width = 0.35\textwidth]{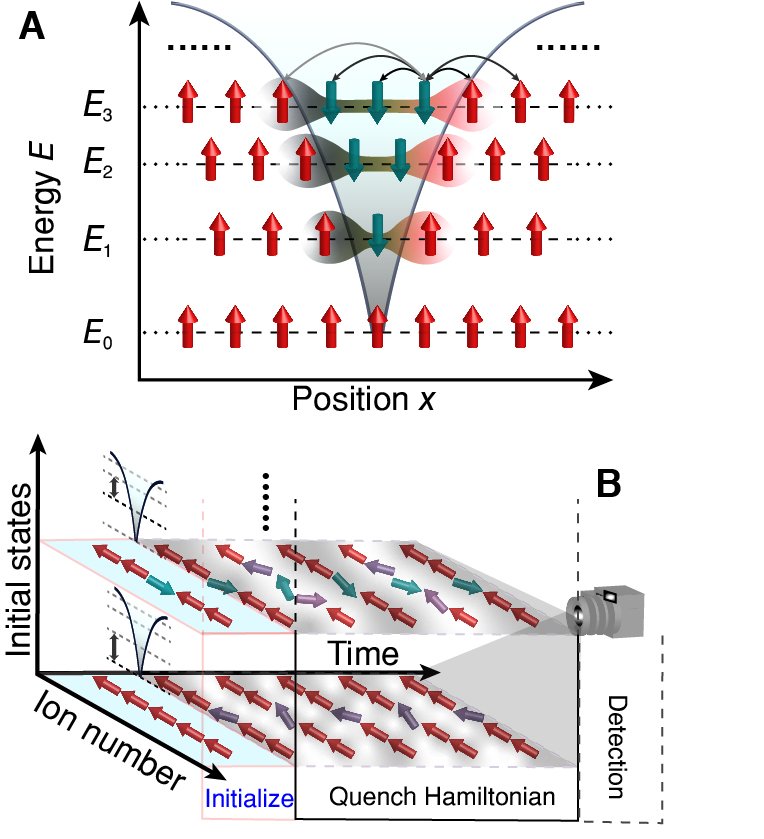}
    \caption{\textbf{Effective confining potential and experiment sequence.} \textbf{(A)} Magnetic domain walls in Ising spin chains can experience an effective confining potential that increases with distance analogously to the strong nuclear force. This potential results in `meson-like' domain wall bound states (labeled $E_1$ to $E_3$) that can dramatically influence the dynamics of the system \cite{Fangli2019,Mazza2019}. \textbf{(B)} This experiment begins by initializing a chain of trapped-ion spins in a product state. We are able to introduce pairs of domain walls by flipping the initial states of chosen spins. The spins evolve according to the quenched Hamiltonian for some time, after which we measure various observables, such as magnetizations of each individual spin along a desired axis.
    }
    \label{fig:ExptCartoon}
\end{figure}

Quantum simulators allow the study of out-of-equilibrium physics of quantum many-body systems in a well-controlled environment \cite{Feynman1982,Georgescu2014}. Here, we use trapped-ion quantum simulators \cite{PorrasCirac2004,Bohnet2016,Jurcevic2017,DPT} to observe real-time domain wall confinement dynamics in a spin chain following a quantum quench, or sudden change in the Hamiltonian (Fig. \ref{fig:ExptCartoon}). We show that confinement can suppress the spreading of correlations even in the absence of disorder, and that quench dynamics can be used to characterize the excitation energies of confined bound states. Additionally, we measure the number of domain walls generated by a global quench, in and out of the confinement regime. Finally, we demonstrate that the number of domain walls can be an effective probe of the transition between two distinct dynamical regimes \cite{DPTHalimeh,Zunkovic2018}.

\renewcommand{\figurename}{\textbf {Fig.}}

Confinement in many-body systems occurs in one of the classic models of statistical mechanics: the Ising spin chain with both transverse and longitudinal magnetic fields. A non-zero longitudinal field confines pairs of originally freely-propagating domain wall quasiparticles into `meson-like' bound states in a short-range interacting system \cite{Kormos2017,James2019,Mazza2019}.
However, recent theoretical efforts \cite{Lerose2019,Fangli2019} have demonstrated that {\emph{long-range}} Ising interactions, instead of an additional longitudinal field, can naturally induce a confining potential between pairs of domain walls (Fig. \ref{fig:ExptCartoon}A).  As a consequence of confinement, the low-energy spectrum of such an Ising system can feature `meson-like' bound domain wall quasiparticles (Fig. \ref{fig:ExptCartoon}A) \cite{Kormos2017,Fangli2019}.

In this report, we use a trapped-ion quantum simulator to investigate confinement in a many-body spin system governed by the Hamiltonian ($\hbar=1$)
\begin{equation}\label{ConfinementH}
    \textit{H} = -\sum_{i<j}^{L} J_{i,j} \sigma_\textit{i}^x\sigma_\textit{j}^x - B \sum_{i}^{L} \sigma_i^z. 
\end{equation}
Here, $\sigma_\textit{i}^\gamma$ ($\gamma = x,y,z$) is the Pauli operator acting on the \textit{i}th spin, $J_{i,j} \approx J_{0}/|i-j|^\alpha$ is the power-law decaying Ising coupling between spins \textit{i} and \textit{j} with tunable exponent $\alpha$, $J_{0} > 0$, \textit{B} is the effective transverse field, and $L$ is the number of spins \cite{Kihwan2009,Pagano2018,Supplementary}. We encode each spin in the ground-state hyperfine levels, $\ket{\uparrow}_z \equiv \ket{F = 1, m_F = 0} $ and $\ket{\downarrow}_z \equiv \ket{F = 0, m_F = 0}$, of the ${}^2S_{1/2}$ manifold of a ${}^{171}\text{Yb}^+$ ion. 
The Ising couplings are produced via spin-dependent optical dipole forces, with power-law exponents $\alpha$ ranging from $0.8$ to $1.1$ and $J_0/2 \pi$ ranging from $0.23$ kHz to $0.66$ kHz \cite{Supplementary}.

To study the real-time dynamics of the spin chain, we use a quantum quench to bring the system out of equilibrium (Fig.  \ref{fig:ExptCartoon}B). 
We first initialize the spins in a product state, polarized either along the $x$ or $z$-directions of the Bloch sphere. Using a tightly focused individual addressing laser \cite{Lee2016}, we are able to prepare domain walls in various initial state configurations (Fig. \ref{fig:ConCorrs}C, F, I). 
After preparing the desired initial state, we perform a sudden quench of the Hamiltonian (\ref{ConfinementH}). Following the time evolution of the system, we use spin-dependent fluorescence to measure the state of each spin. From this data, we calculate the time-evolution of magnetizations, $\braket{\sigma^x_i(t)}$ or $\braket{\sigma^z_i(t)}$, and connected correlations 
\begin{equation}
C_{i,j}^x(t)= \braket{\sigma^x_i(t) \sigma^x_j(t)}-\braket{\sigma^x_i(t)}\braket{\sigma^x_j(t)}.
\end{equation}
No post-processing or state preparation and measurement (SPAM) correction has been applied to any of the data reported below.

\renewcommand{\thefigure}{2}
\begin{figure*}[tb]
    \centering
    \includegraphics[scale=1.0]{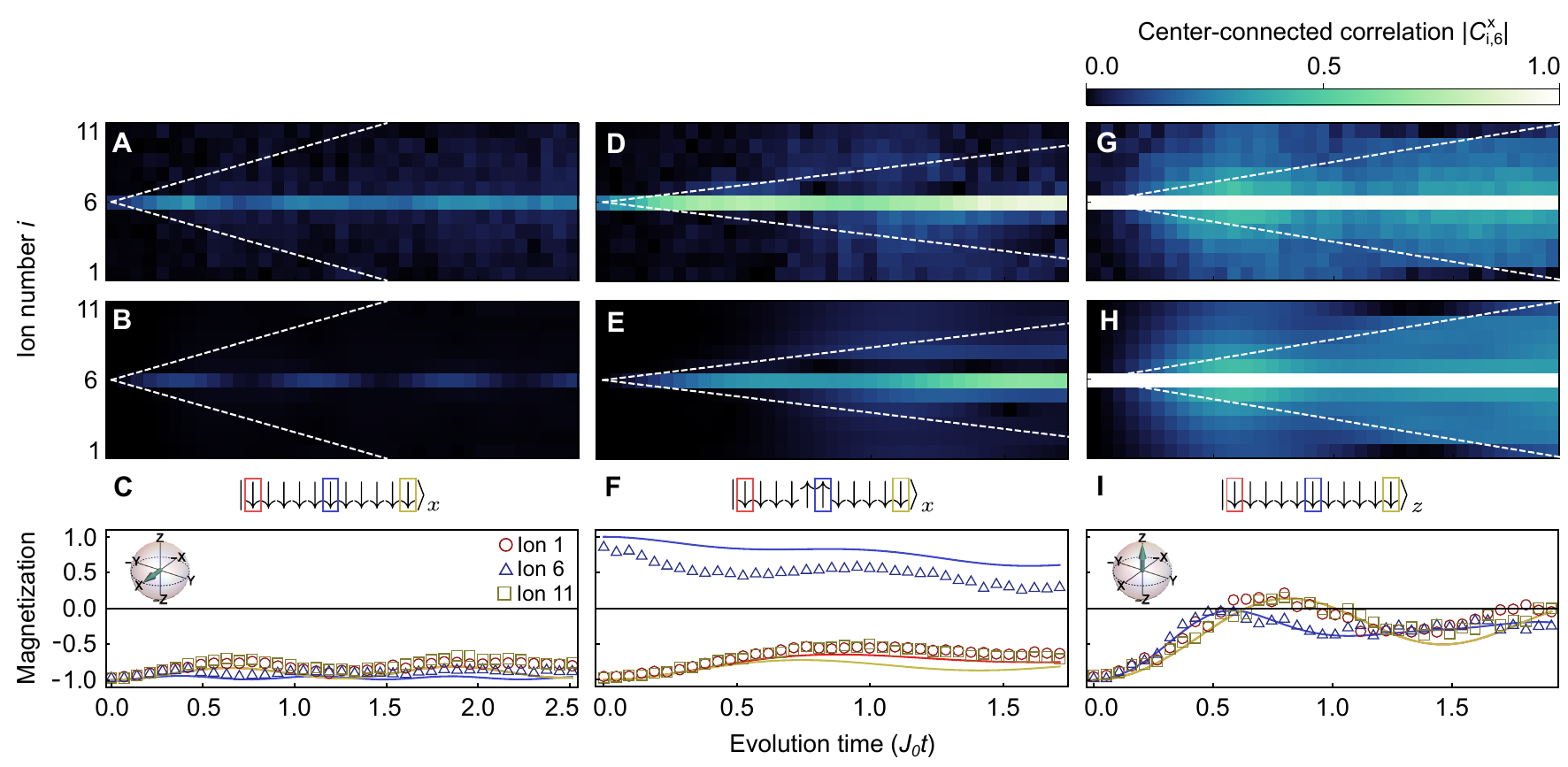}
    \caption{ \textbf{Confinement Dynamics at }$\boldsymbol{ B/J_{0} \approx 0.75, L=11}$. The top row shows the absolute value of experimental center-connected correlations $|C_{i,6}^x(t)|$ averaged over 2000 experiments. The middle row shows $|C_{i,6}^x(t)|$ calculated by solving the Schr\"{o}dinger equation. Expected correlation propagation bounds, or light cones, in the $\alpha \rightarrow \infty$ limit are represented by dashed white lines. The bottom row shows measured individual-spin magnetizations along their initialization axes, $\braket{\sigma^{x,z}_i(t)}$, averaged over 2000 experiments (400 experiments for \textbf{(I)}). Symbols represent magnetization data and solid colored curves represent theoretical magnetizations calculated by solving the Schr\"{o}dinger equation. All magnetization error bars, $\pm 1\sigma$, are smaller than their plot symbols and are not shown.
    \textbf{(A,B,C)} show a low-energy initial state containing zero domain walls. Individual magnetizations are $\braket{\sigma^x_i(t)}$.
    \textbf{(D,E,F)} show a low-energy initial state containing two domain walls, with a center domain of two spins. Individual magnetizations are $\braket{\sigma^x_i(t)}$. We attribute the discrepancy between the experimental magnetization data and numerics to imperfect state initialization.
    \textbf{(G,H,I)} show a high-energy initial state containing many domain walls. \textbf{(I)} Individual magnetizations are $\braket{\sigma^z_i(t)}$.
    }
    \label{fig:ConCorrs}
\end{figure*}

\renewcommand{\thefigure}{3}
\begin{figure*}[h!t]
    \includegraphics[scale=1.0]{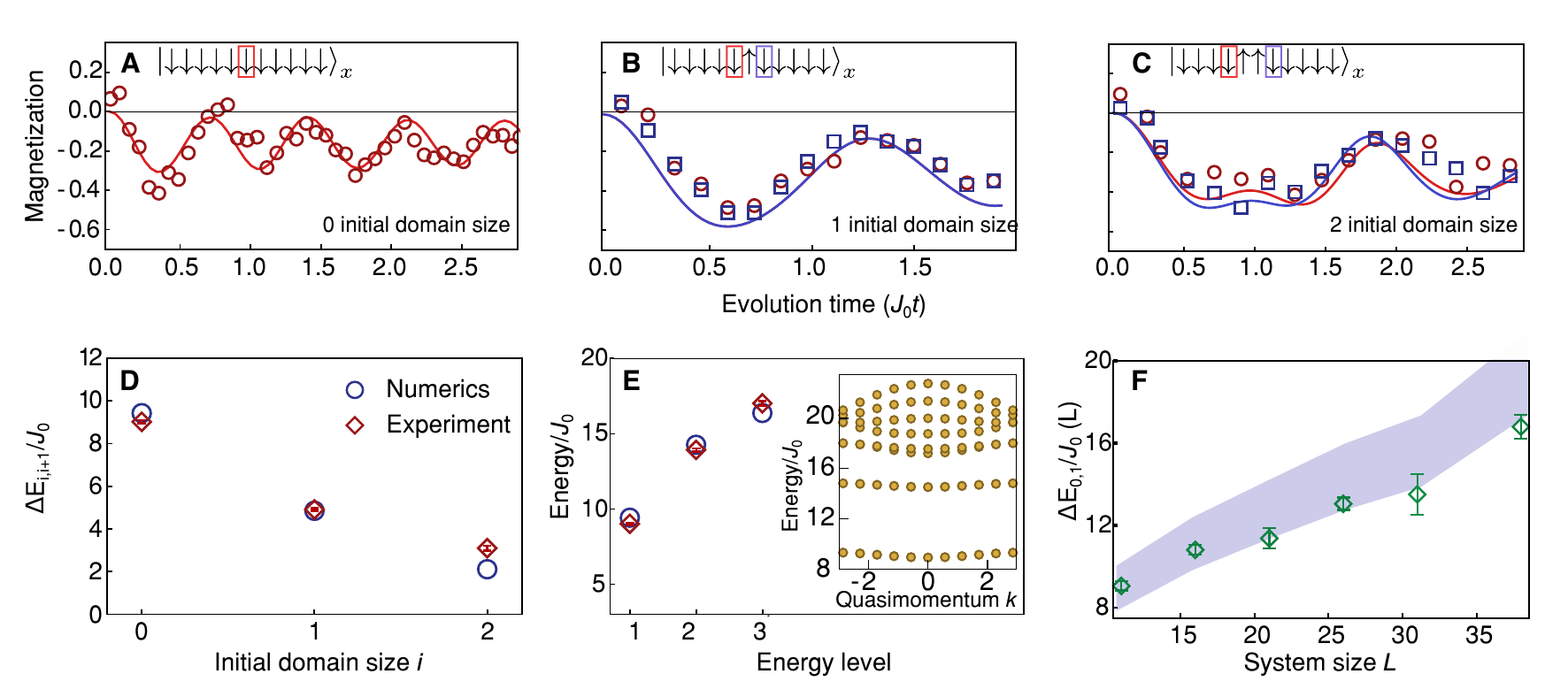}
    \caption{\textbf{Low-Energy Excited States.}
    (\textbf{A-C}) show the magnetizations of the boxed spins on the edges of the center domain at $B/J_{0} \approx 0.75$. These magnetization oscillation frequencies correspond to the normalized energy gap, $\Delta E_{i, i+1}/J_0$, between the prepared state $i$ and the adjacent higher-energy bound state $i+1$. Solid colored lines represent theoretical calculations of dynamics by solving the Schr\"{o}dinger equation. The error bars, $\pm 1\sigma$, are smaller than their plot markers and are not shown in A-C. (\textbf{A}) Zero initial domain size: $\Delta E_{0, 1}/J_0 $ is given by the frequency of the 6th spin. (\textbf{B}) Initial domain size of one: $\Delta E_{1, 2}/J_0$ is given by the frequency of  the 5th and 7th spins. (\textbf{C}) Initial domain size of two: $\Delta E_{2, 3}/J_0$ is given by the frequency of  the 4th  and 7th spins. (\textbf{D}) $\Delta E_{i, i+1}/J_0$ for $i\leq 2$ are measured with three different initial domain size spin configurations at  $B/J_{0} \approx 0.75$. The first three energy gaps ($i\leq2$) are extracted from the magnetization oscillation frequencies shown in the top row. 
    (\textbf{E}) We construct the bound state energy levels at quasimomentum $k \approx 0$ using experimental data in D, where $E_0/J_0$ is set to be zero. Inset: Theoretical bound state energy bands with different quasimomentum, $k$, within the two domain wall model \cite{Fangli2019,Supplementary}.  (\textbf{F}) Scaling of $\Delta E_{0, 1}/J_0$ with system size at $B/J_0 \approx 1$. The blue shaded region shows the two domain wall model \cite{Fangli2019,Supplementary} numerical prediction of $\Delta E_{0, 1}/J_0$, with a confidence band considering $\pm 10\%$ fluctuations in the Ising interaction strength $J_0$. }  
    \label{Fig3}
\end{figure*}

To understand the effect of confinement on information spreading, we measure the absolute value of connected correlations along $x$, the Ising direction (Fig. \ref{fig:ConCorrs}). When the initial state contains a small number of domain walls, correlations spread with a considerably smaller velocity than the predicted nearest-neighbor interacting limit ($v_{0}=4 B$ \cite{Kormos2017}, Fig. \ref{fig:ConCorrs}). While correlation functions typically exhibit a light cone behavior following a quantum quench \cite{Calabrese2006,Cheneau2012,Eisert2015}, we observe strongly suppressed spreading and localized correlations throughout the evolution \cite{James2019}. This indicates that confinement, induced by long-range interactions, localizes pairs of domain walls at their initial conditions \cite{Supplementary}.

In stark contrast, we find that correlations exhibit superballistic spreading, despite quenching under the same Hamiltonian, in the case of the initial state polarized in the transverse direction $z$ (Fig. \ref{fig:ConCorrs}).
In this case, the initial state is a linear superposition of all possible spin configurations in the $x$-direction, and thus contains a large number of domain walls. Unlike the previous initial states, this initial state has an energy density relatively far from the bottom of the many-body spectrum. The long-range interactions among these domain walls lead to fast relaxation and quantum information spreading. These results imply that this confinement effect has a significant impact only on the low-energy excitations of the system, which is consistent with recent theoretical studies \cite{Kormos2017,Lerose2019,James2019,Fangli2019,Mazza2019}. 

To observe the effect of confinement on the thermalization of local observables, we  measure the relaxation of magnetizations for the above initial states and compare with numerical predictions (third row of Fig. \ref{fig:ConCorrs}). We see that, for the low-energy states, local magnetizations retain long memories of their initial configuration and exhibit slow relaxation (Fig. \ref{fig:ConCorrs}C, F). Conversely, for the high-energy initial state, local magnetizations quickly relax to their thermal expectation values (Fig. \ref{fig:ConCorrs}I). This is consistent with the observation that correlations quickly distribute across the entire system ({Fig. \ref{fig:ConCorrs}H}). We emphasize that the observed slow thermalization is a consequence of confinement, distinct from many-body localization with quenched disorder \cite{Schreiber2015,Nandkishore2015,Hess2017}. 

In order to quantitatively probe excitation energies of bound domain wall states, we prepare initial states polarized along the $x$-direction and vary the number of spins separating the two initial domain walls (insets of Fig. \ref{Fig3}A-C). Then, we quench the Hamiltonian (\ref{ConfinementH}) and measure the time-evolution of local magnetizations along the transverse direction, $\braket{\sigma^z_i (t)}$. In the confined regime, all local observables should exhibit oscillations with frequencies proportional to the energy gap between adjacent bound states before thermalizing \cite{Kormos2017,Fangli2019}. Here, we choose a single-body spin observable, $\braket{\sigma^z_i (t)}$, at the center of the chain (for 0 initial domain walls) or at the outer boundary of the initial domain (for 2 initial domain walls). 
We make this particular choice in order to minimize edge effects from the finite spin chain and maximize the matrix elements of this observable between two adjacent states (Fig. \ref{fig:ExptCartoon}A).

Following this prescription, we extract oscillation frequencies using single-frequency sinusoidal fits of $\langle \sigma_\textit{i}^z(t) \rangle$ to obtain the energy gap between each initialized state and the neighboring excited state (Fig. \ref{Fig3}A-C).
We compare these extracted energies to values predicted by numerical simulation \cite{Supplementary}. We find excellent agreement between the measured energies and the energies predicted numerically (Fig. \ref{Fig3}E). Using these experimentally measured energy gaps, we can systematically construct the low-energy excitation spectrum of the many-body system for quasimomentum $k\approx0$ (Fig. \ref{Fig3}E). In general, quasiparticles with arbitrary quasimomenta can be excited by a quantum quench. However, since the confining potential is steep, excited quasiparticles remain localized and their quasimomenta are close to zero. Furthermore, leveraging the scalability of trapped-ion systems, we perform this experiment with up to 38 spins. In order to numerically investigate these large system sizes, we use a phenomenological two domain wall model \cite{Fangli2019,Supplementary}. With this model, by restricting the full Hilbert space to a subspace of states containing only zero or two domain walls, we are able to calculate the bound quasiparticle spectrum of Hamiltonian (\ref{ConfinementH}) for classically-intractable system sizes (Fig. \ref{Fig3}F). We find reasonable agreement in the first excitation energy gap, $\Delta E_{0, 1}$, between the experimental data and numerical predictions for all system sizes (Fig. \ref{Fig3}F). We attribute the systematic discrepancy in larger systems to variations in $J_0$ during the time evolution \cite{Supplementary}. These results, taken together, suggest that quench dynamics are dominated by the confinement effect between two domain wall quasiparticles.

\renewcommand{\thefigure}{4}
\begin{figure*}[h!tb]
    \includegraphics[scale =1]{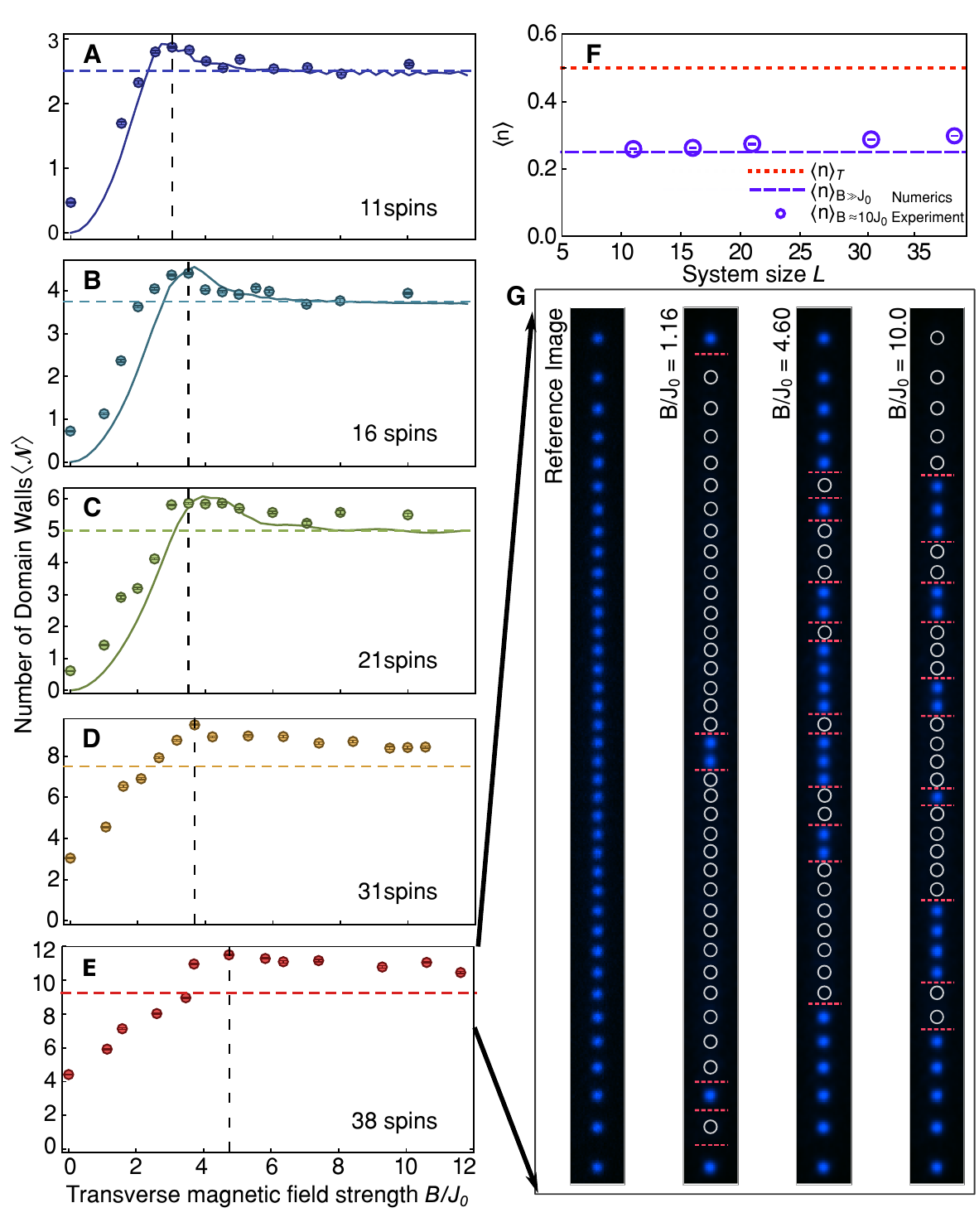}
    \caption{{{\bf{Number of Domain Walls in Two Dynamical Regimes}}. (\textbf{A-E}) Exploring the two dynamical regimes with increasing transverse $B$-field strength in different system sizes. Circular dots indicate experimental data. Horizontal lines show theoretical predictions of $\langle\textit{$\mathcal{N}$}\rangle$ =  $0.25(L-1)$ at $B\gg J_0$.  Colored continuous lines represent numerical results predicted by solving  the Schr\"{o}dinger equation. Vertical dashed lines indicate the experimental maxima of $\langle\textit{$\mathcal{N}$}\rangle$. (\textbf{F}) Dashed purple horizontal line shows the theoretical prediction of domain wall density at $B \gg J_0$, $\langle\textit{n}\rangle_{B\gg J_0}$. The purple circular dots indicate experimental data of domain wall density at $B \approx 10 J_0$, $\langle\textit{n}\rangle_{B \approx 10 J_0}$. The dashed red line at $\langle\textit{n}\rangle_T = 0.5$ shows the density of domain walls using the canonical ensemble at infinite temperature. All the experimental data is integrated within the time interval $J_0 t_1 \approx 0.34$ and $J_0 t_2 \approx 0.73$. (\textbf{G}) Reconstructed images based on binary detection of spin states. The leftmost image is a reference of a 38 ion chain in a `bright' state ($\ket{\uparrow}_x$). At the beginning of the experiment, the spins are  initialized in the `dark' state ($\ket{\downarrow}_x$). The three right images show experimental data of a combination of `bright' and `dark' states, marked in blue and white circles respectively, for three different $B/J_0$ values within the integrated time frame. The occurrences of domain walls are highlighted with orange horizontal dashed lines. 
    }
    }
    \label{Fig4}
\end{figure*}

We now go beyond the confinement regime to study the number of domain walls generated by the quantum quench for a wide range of transverse $B$-field strengths. 
Although we still prepare an initial state polarized along $\ket{\downarrow}_x$, for large $B$, the strong quench can excite a large number of domain walls which are no longer bounded. We thus expect that the out-of-equilibrium dynamics are no longer captured by the confinement picture for these parameters. To explore this regime, we measure the cumulative time average of the total number of domain walls, 

\begin{equation}\label{Kinks}
    \langle\textit{$\mathcal{N}$}\rangle =\frac{1}{t_{2}-t_{1}}\int_{t_{1}}^{t_{2}} \sum_{i=1}^{L - 1} \frac{\langle 1- \sigma_\textit{i}^x(t) \sigma_\textit{i+1}^x(t)\rangle}{2},
\end{equation}
where $t_1$ and $t_2$ enclose a window where $\langle\textit{$\mathcal{N}$}\rangle$ converges to a stable value \cite{Supplementary}. We measure $\braket{\mathcal{N}}$ as a function of $B$ for different system sizes (Fig. \ref{Fig4}A-E). We observe that, for small $B$ fields, Ising interactions dominate the dynamics and the global quench can only excite a small number of domain walls. However, for a large enough transverse field, the number of generated domain walls saturates to a value that scales nearly linearly with system size (Fig. \ref{Fig4}). Here, we observe a transition between these two dynamical regimes at intermediate values of $B$ for different system sizes. To illustrate the population of domain walls in different regimes, we show typical single-shot images of the quenched state of 38 ions for different transverse $B$-fields in Fig. \ref{Fig4}G. We indeed see that a small (large) number of domain walls is generated by the quench with small (large) $B$ field. Although the numerical simulation of the exact dynamics of the largest system size of 38 spins can be challenging, we can intuitively understand the distinguishing behaviors. When we increase $B$ to values significantly larger than $J_0$, all spins undergo Larmor precession around the $z$-axis of the Bloch sphere, which allows us to predict that $\braket{\mathcal{N}}$ saturates to $0.25  (L-1)$  when $B\rightarrow \infty$ \cite{Supplementary,Calabrese2012}. We note that, for $B \gg J_0$, the experiment operates in the prethermal region in which a transient Hamiltonian is approximately conserved for an exponentially long time  \cite{Abanin2017,HalimehPrethermal2017,Tran2019,Machado2019}. Therefore, we expect the number of domain walls to approach the thermal value, $ \braket{n}_T = 0.5$, only after an exponentially long time, beyond the reach of this experiment. The experimental results agree with the numerical prediction for system sizes within the reach of numerical simulations. We attribute the discrepancies at large system sizes to bit-flip events due to detection errors and off-resonant coupling to motional degrees of freedom \cite{Supplementary}, and to finite effective magnetic fields $B$ compared to the total interaction energy \cite{Essler2014}, that is increasing with system size due to its long-range character.

In summary, we have presented a real-time observation of domain wall confinement caused by long-range interactions in trapped-ion spin systems. By measuring oscillating magnetizations, we were able to construct the spectrum of low-energy domain wall bound states. Furthermore, we observed a transition between distinct dynamical behaviors using the number of domain walls generated by the global quench. This work demonstrates that confinement, naturally induced by long-range interactions, may provide a novel mechanism for protecting quantum information without engineering disorder. Such a feature may be applied in future studies to use long-range interactions to stabilize non-equilibrium phases of matter. All together, this work establishes the utility of trapped-ion quantum simulators for precisely studying real-time dynamics of many-body systems, potentially extending to exotic phenomena such as quark collision and string breaking \cite{Verdel2019}.

\section*{Acknowledgments}
We are grateful to D. A. Abanin, P. Bienias, P. Calabrese, M. Dalmonte, A. Deshpande, A. Gambassi, M. Heyl, A. Lerose, J. Preskill, A. Silva, P. Titum, and R. Verdel for enlightening discussions. 
\textbf{Funding: }This work  is supported by the NSF PFCQC STAQ program, the AFOSR MURIs on Quantum Measurement/Verification, the ARO MURI on Modular Quantum Systems, the AFOSR and ARO QIS and AMO Programs, the DARPA DRINQS program, the DOE BES award DE-SC0019449, the DOE HEP award DE-SC0019380, the NSF QIS program, and the NSF Physics Frontier Center at JQI.

\noindent\textbf{Author Contributions: }
W.L.T. and P.B. contributed equally to this work.
F.L. and G.P. suggested the research topic. W.L.T., P.B., G.P., K.S.C., A.D., L.F., H.B.K., A.K., W.M., and C.M. all contributed to experimental design, construction, data collection, and analysis. F.L., R.L., S.W., K.S.C., W.L.T., P.B., and G.P. carried out numerical simulations. F.L., R.L., S.W. and A.V.G. provided theoretical support. All authors contributed to the discussion of the results and the manuscript.

\noindent\textbf{Competing Interests: }
The authors declare competing financial interests: C.M. is Co-Founder and Chief Scientist at IonQ, Inc.

\noindent\textbf{Data and materials availability: }
The data presented in the figures of this report are available from the corresponding authors upon reasonable request.

\section{Supplementary Materials}

\subsection{Trapped-ion Quantum Simulators}
In this work, we employ two quantum simulators, which we refer to as System 1 \cite{Kihwan2009} and System 2 \cite{Pagano2018}. System 1 is a room temperature trapped-ion apparatus. It employs a 3-layer linear Paul trap with transverse center-of-mass (COM) motional frequency $\nu_{COM} = 4.7 $ MHz and axial COM frequency $\nu_{z} \approx 0.5 $ MHz \cite{Kihwan2009}. The main limitation of this apparatus is the rate of collisions with the residual background gas in ultra-high vacuum (UHV), limiting the practical size of the chain. During such collision events, the ion crystal melts and ions are ejected from the trap due to RF-heating. However, this apparatus has individual addressing capabilities, allowing for initialization of arbitrary spin flips, which is crucial in this work. Therefore, we investigate low-energy domain wall bound states in smaller system sizes with this apparatus.

System 2 is a linear blade Paul trap in a cryogenic environment with only global qubit control \cite{Pagano2018}. The trap is held at $\approx$ 8 K in a closed cycle cryostat, where the background pressure is below $10^{-12}$ Torr due to differential cryopumping. This allows for longer storage lifetimes of large ion chains as compared to System 1.  For this reason, System 2 can support larger chains to measure the lowest bound state energy and investigate the two distinct dynamical regimes by increasing the transverse $B$-field. To take the anharmonicity of the trap into account, we measure all the transverse motional modes of the ion chain. The transverse motional frequencies are $\nu_{COM}^{x} = 4.4 $ MHz and $\nu_{COM}^{y} = 4.3$ MHz, the $x$-tilt frequency $\nu^x_{tilt}$ ranges from 4.37 MHz to 4.38 MHz and the $y$-tilt frequency $\nu^y_{tilt}$ ranges from 4.24 MHz to 4.25 MHz and, depending on the number of trapped ions. 

\subsection{Initial State Preparation}
In both systems, every experiment begins by Doppler cooling a chain of trapped ${}^{171} \text{Yb}^+$ ions using 369.5 nm light red-detuned from the ${}^2S_{1/2}$ to ${}^2P_{1/2}$ transition. The ions are initialized to the $\ket{\downarrow}_z$ qubit state, defined as the ${}^2S_{1/2} \ket{F = 0, m_F = 0}$ hyperfine level, by an incoherent optical pumping process. Optical pumping takes roughly $20 ~\mu$s and initializes all ions with at least 99\% fidelity. Next, the ions are cooled to their motional ground state ($\leq 0.1$ average motional quanta for the COM mode) with Raman sideband cooling.
Once the spins are cooled and initialized, we may prepare them in product states along any axis of the Bloch sphere by applying global rotation pulses. System 1 has the ability to manipulate spins with an individual addressing beam focused to a waist of $500$ nm, 3 to 4 times smaller than the typical inter-ion spacing in System 1. This beam applies a fourth-order ac Stark shift to the qubit splitting, causing an effective $\sigma^z_i$ rotation on a single spin \cite{Lee2016}. This rotation can be mapped to a rotation about any axis with global $\pi/2$-pulses, which allows preparation of product states with arbitrary spin flips.

\subsection{State Detection}
Following an experiment, we measure each spin's magnetization with spin-dependent fluorescence  using an  Andor  iXon  Ultra  897  EMCCD  camera which resolves the magnetizations of individual spins. A 369.5 nm laser resonant with the ${}^2S_{1/2}\ket{F=1}$ to ${}^2P_{1/2}\ket{F=0}$ transition (linewidth $\gamma/2\pi \approx 19.6$ MHz) causes photons to scatter off each ion if the qubit is projected to the $\ket{\uparrow}_z$ state. Conversely, ions projected to the $\ket{\downarrow}_z$ qubit state scatter a negligible number of photons because the laser is detuned from resonance by the ${}^2S_{1/2}$ hyperfine splitting. By applying global $\pi/2$-pulses, we are able to rotate the $x$ and $y$ bases into the $z$ basis. This allows us to measure all individual magnetizations and many-body correlators along any single axis in a single shot.

Both systems collect scattered 369.5 nm photons using a finite conjugate 0.4 NA objective with total magnification of 45x for System 1 and 90x for System 2. Before taking data, high-contrast calibration images of the ion chain, illuminated by Doppler cooling light, are used to identify a region of interest (ROI) on the camera sensor for each ion. System 2 may take multiple calibration images in between experiments to account for slow drifts of the ions' positions. During data collection, System 1 (2) integrates collected fluorescence for 0.65 (1.0) ms, after which a pre-calibrated binary threshold is applied to discriminate the qubit state of each ion with approximately 97\% accuracy per ion. The dominant detection error sources are: off-resonant mixing of qubit states during the detection period, cross-talk between ion ROIs due to small inter-ion spacings near the center of the chain, electronic camera noise, and laser power fluctuations. We do not perform any post-processing, including state preparation and measurement (SPAM) correction, on the data presented in this report.

\subsection{Generating the Ising Hamiltonian}
We generate spin-spin interactions by applying spin-dependent dipole forces with a pair of non-copropagating 355 nm Raman beams for which the beatnote wavevector, $\Delta k$, is aligned along the transverse motional modes of the ion chain. These two beams are controlled with acousto-optic modulators that generate a pair of beat note frequencies $\nu_0 \pm \mu$ for the M\o{}lmer S\o{}rensen (MS) scheme \cite{Molmer1999}. In the Lamb-Dicke regime \cite{NistBible}, the laser-ion interaction gives rise to an effective spin-spin Hamiltonian where the coupling between spins $i$ and $j$ is:
\begin{equation}\label{JijInteraction}
    J_{i,j} = \Omega^2 \nu_R \sum_{m}\frac{b_{i,m}  b_{j,m} }{\mu^2-\nu^2_m}  \approx \frac{J_0}{|i-j|^\alpha}
\end{equation}
where $\Omega$ is the global Rabi frequency, $\nu_R = \hbar {\Delta k}^2/(2 M)$ is the recoil frequency, $\nu_m$ is the frequency of the $m$-th motional mode, $b_{im}$ is the eigenvector matrix element of the $i$-th ion's participation in the $m$-th motional mode ($\sum_i|b_{im}|^2=\sum_m|b_{im}|^2 = 1$), and $M$ is the mass of a single ion. 

Unlike System 1, where $\Delta k$ is aligned along one set of transverse motional modes, System 2 couples to both sets of transverse motional modes as the Raman beams project onto the two radial principal axes of the trap.  While coupling to these additional modes creates the same Hamiltonian as System 1  (\ref{ConfinementH}), the coupling strengths between ions may differ. To account for this, (\ref{JijInteraction}) can be generalized to:
\begin{align}\label{JijInteractionTwoModes}
    J_{i,j} &= J_{i,j}^x+ J_{i,j}^y \\
    J_{i,j}^\beta &=  \Omega^2_\beta \nu_R^\beta  \sum_{m}\frac{b_{i,m}^\beta   b_{j,m}^\beta  }{\mu^2-(\nu^\beta)^2}, ~~ \beta = x,y 
\end{align}
where $\nu_R^\beta$ is the recoil frequency given by the $\beta$ projection of $\Delta k$ ($\Delta k^x$ and $\Delta k^y$ ). Both experiments work in the MS regime where the beatnote frequencies are detuned by $\mu$ far from all the motional sidebands, $|\mu-\nu_m| 	\gg \eta \Omega$, where $\eta$ is the Lamb-Dicke parameter, to suppress phonon production via virtually coupling spins to motion.

The approximate power law exponent, $\alpha$, in (\ref{JijInteraction}) theoretically can be tuned within the range $0<\alpha<3$. However, in practice, we are restricted to $0.5<\alpha<1.8 $ to avoid motional decoherence and experimental drifts. Therefore, in this work, we are in the regime where all excitations within the two domain wall model are bounded, where $\alpha<2$ (see Section 1.5 for details). In the reported experiments, the power-law exponent is $\alpha = 1.1$ with $J_{0}/2\pi$ ranging from $0.45$ kHz to $0.66$ kHz for System 1. System 2 operates in the regime with $\alpha$ between 0.8 and 1 with $J_0/2\pi$ ranging from $0.23$ kHz to $0.42$ kHz. 

We apply a global offset to the two Raman lasers by $2B_z$, creating a rotating frame shift between the qubit and the Raman beatnote to generate an effective transverse magnetic field $B_z$. We limit the effective transverse $B$-field to $B \ll \eta \Omega_{COM} \ll \delta_{COM}$, where $\eta \Omega_{COM}$ is the COM sideband rabi frequency and $\delta_{COM}$ is the beatnote's detuning away from the transverse COM mode.

These trapped-ion quantum simulators natively realize an antiferromagnetic Ising model. All measured observables $O(t)$ of the evolution are real and symmetric under time-reversal. This implies the measured observables of Hamiltonians $\textit{H}$ and $\textit{-H}$ are the same. Therefore, the expectation values we obtain from $J_{i,j}>0$ and $B>0$ are identical to $J_{i,j}<0$ and $B<0$. For this reason, we are able to simulate the dynamics of a ferromagnetic system \cite{Schachenmayer2013}. 

\subsection{Two Domain Wall Model}
Previous experimental and theoretical studies \cite{Coldea2010,Fangli2019} have found that the low-energy excitations of confining Hamiltonians, such as Eq. (\ref{ConfinementH}), largely consist of states containing zero or two domain walls. By restricting the Hilbert space to include only these states, we can build a relatively simple phenomenological model that mimics the low-energy behavior of the system. Liu et al. describes such a `two-kink model' for a ferromagnetic long-range transverse field Ising chain with closed boundary conditions and $B<J_0$ in \cite{Fangli2019}, which we will summarize here.

The Hilbert space of this model contains states with two down-aligned domains surrounding an up-aligned domain of length $l$. These domains are separated by two domain walls: one between spin positions $j-1$ and $j$ and another between positions $j+l-1$ and $j+l$. Such a state $\ket{j,l}$ has the form
\begin{equation}
\ket{j,l}=\ket{\downarrow ... \downarrow \downarrow_{j-1} \uparrow_{j} \uparrow ... \uparrow \uparrow_{j+l-1} \downarrow_{j+l} \downarrow ... \downarrow}.
\end{equation}

The Hamiltonian for this set of basis states is given by Eq. (2) in \cite{Fangli2019}. For a translational-invariant system, it is useful to transform to a set of quasimomentum basis states $\ket{k,l}=(1/L)\sum^L_{j=1}\text{exp}(-ikj - ikl/2)\ket{j,l}$, which are eigenstates of the Hamiltonian

\begin{align}\label{twokinkmodelHamiltonian}
\begin{split}
H=\sum_{k,l} &V(l)\ket{k,l}\bra{k,l} - 2B \text{cos}\left(\frac{k}{2}\right)\ket{k,l}\bra{k,l+1}\\ &-2B\text{cos}\left(\frac{k}{2}\right)\ket{k,l}\bra{k,l-1}.
\end{split}
\end{align}
Both terms involving the transverse field $B$ describe the effective kinetic energy of the domain walls with quasimomentum $k$. The potential $V(l)$ depends on the interaction strengths $J_{i, j}$ in the system
\begin{equation}
    V(l)= - \sum_{i<j}^{L} J_{i, j} s_i(\mathcal{S}) s_j(\mathcal{S})
\end{equation}
where $s_i(\mathcal{S}) = \pm1$ is the value of the spin at site $i$ corresponding to the configuration $\mathcal{S}$ with domain of length $l$. This Hamiltonian can be diagonalized to reveal the presence of energy bands in the low-energy spectrum (inset of Fig. \ref{Fig3}E). These bands represent domain wall states bounded by the potential $V(l)$. For $\alpha<2$ this potential is unbounded and all domain wall pairs will be confined into quasiparticles.

The trapped-ion spin system is finite with open boundary conditions. To minimize deviations from this model due to finite-size effects, we consider only those states with short, up-aligned domains ($l \ll L$) centered in the spin chain. With this constraint, we find good agreement between exact diagonalization ($L \leq  21$), the two domain wall model, and experimental results. The two domain wall model numerics for this experiment are implemented by taking the experimental $J_{i,j}$ matrix to calculate the energy gaps for each experiment.  We first extract a vector of interaction parameters from the experimental interaction matrix, $J_{k,j}$ by fixing the site $k$ to be the center ion. Then, we theoretically put the ions on a ring and impose a periodic boundary condition by requiring the Ising interaction to be translationally invariant, i.e. $J_{l, m}= J_{k, k+m-l}$. Using this method,  we are able to obtain the spectrum of energy bands and energy gaps for  the trapped-ion system by diagonalizing Eq. (\ref{twokinkmodelHamiltonian}) (Fig. \ref{Fig3}E).

\subsection{Domain Wall Localization in the Confinement Regime}
In the main text, we claim that slow or negligible spreading of correlations following a quench of the confining Hamiltonian (\ref{ConfinementH}) indicates that domain walls are localized at their initial positions. In this section, we extend that argument by measuring the average number of domain walls at each available position of an $L=11$ spins chain after a quench. The average number of domain walls $\braket{N_j(t)}$ at site $j \in \{ 1,L-1 \}$ at time $t$ is given by 
\begin{equation}
    \braket{N_j(t)}= \frac{\braket{1-\sigma_{j}^x(t) \sigma_{j+1}^x(t)}}{2}.
    \label{kinkevolution}
\end{equation}
Fig. \ref{fig:KinkLocalization} shows both experimental measurements and numerics of the evolution of $\braket{N_j(t)}$ for six initial states. The first three rows correspond to data shown in Figs. \ref{fig:ConCorrs} and \ref{Fig3}A-E and represent states within the two domain wall model. In these cases, pairs of domain walls are strongly localized near their initial positions, showing excellent agreement with numerics. The bottom two rows show higher-energy initial states outside of the two domain wall regime. The N\'{e}el (staggered) state along $x$ is initialized with domain walls at every position, while each site in the $z$-polarized state is initialized with, on average, one half of a domain wall. These high energy density states are not expected to show domain wall confinement.

\subsection{Domain Wall Convergence at High Transverse $B$-field}
In this domain wall investigation, we use the following Bloch sphere mapping: $z \leftrightarrow x$.
The orientation of the $i$th spin in the Bloch sphere is defined as $|\psi_i \rangle = \cos \theta/2 |0\rangle + e^{i\phi}\sin \theta/2 |1\rangle$. Let $|\psi \rangle = |\psi_i\rangle \otimes |\psi_{i+1}\rangle$ since we are interested in a two-body correlator for $\langle\textit{$\mathcal{N}$}\rangle $. At high transverse $B$-field, global Larmor precession about the transverse direction dominates over the Ising interaction term in (\ref{ConfinementH}). The expectation value of the two-body correlator along z is $\langle \sigma_z \sigma_z \rangle = 1-(1-\cos \theta)/2$. Inserting $\langle \sigma_z \sigma_z \rangle$ into Eq. (\ref{Kinks}) gives us 
\begin{equation}
 \braket{\mathcal{N}}=\frac{1}{t_2 - t_1} \int_{t_{1}}^{t_{2}} \sum_{i}^{L-1} \frac{1-(1-\cos \theta(t))/2}{2} dt
\end{equation}
Therefore, $\braket{\mathcal{N}} = 0.25(L-1)$ when $B\gg J_0$.

\subsection{Error Sources}
Experimental noise sources affect the fidelity of the quantum simulation. Many significant noise sources affect the simulation in the form of a `bit-flip error'. A source of bit-flip error is spin-motion entanglement due to off-resonant excitation of the ion chain's motional modes \cite{Wang2012} in the MS regime, where both quantum simulators operate. Unwanted bit-flip errors occur when spin-entangled motional degrees of freedom are traced out at the end of the experiment. The probability of this error to occur on the $i$th ion is proportional to $p_i \approx \sum^N_{m=1} (\eta_{im} \Omega/ \delta_m)^2$, where $\eta_{im} = b_{im}\sqrt{\nu_R/\nu_{m}}$ and  $\delta_{m} = \mu-\nu_m$ is the beatnote detuning from the $m$th motional mode \cite{YukaiThesis}. To minimize this error, we choose $\delta_{COM}$ such that $(\eta_{COM}\Omega / \delta_{COM})^2 \lesssim 1/9$.  Another source of bit-flip error is imperfect state detection (refer to section 1.3. State Detection).
These sources of bit-flip error are typically independent, and therefore will add up in quadrature. We find that, by including the bit-flip error in the $L=11$ spins numerical calculation for $\langle\textit{$\mathcal{N}$}\rangle$, the experimental data agrees well with the error included numerical calculation at $B/J_0 = 0$, as shown in Fig. \ref{fig:BitFlip}. Presently, we are limited to computing this error for $L<15$ spins.

Slow experimental drifts that involve laser light intensity noise at the ions and drifts of the trap frequency (which determines transverse motional modes), over the course of a few hours during data taking, will cause us to average over different effective Hamiltonians. Furthermore, this system has a residual effective linear magnetic field gradient across the ions chain due to the fourth-order Stark shift gradient from imperfect overlap of the two Raman beatnotes. This gradient noise is more prominent for small $B$-fields, causing an effective depolarization of the initial state in the Fig. \ref{fig:DomainWall} data. This effective magnetic gradient is typically $<15$ Hz/$\mu$m across the ion chain. We find that errors caused by this effective magnetic field gradient are much smaller than those caused by bit-flip errors.

Another source of noise is off-resonant Raman scattering during the quantum quench. This error rate is estimated to be $7 \times 10^{- 5}$ Hz per ion, given reasonable experimental parameters. Small errors due to RF heating of the transverse COM motional mode are present in System 1. Although System 2 is in a cryogenic setup that is less susceptible to RF heating, it has mechanical vibrations at 41 Hz and 39 Hz due to residual mechanical coupling to the cryostat \cite{Pagano2018}. This mechanical vibration noise is equivalent to phase-noise on the Raman beams, which leads to qubit dephasing. Therefore, we integrate the number of domain walls before the dephasing occurs (Fig \ref{Fig4}).

\renewcommand{\thefigure}{S1}
\begin{figure*}[t]
    \includegraphics[scale=1.0]{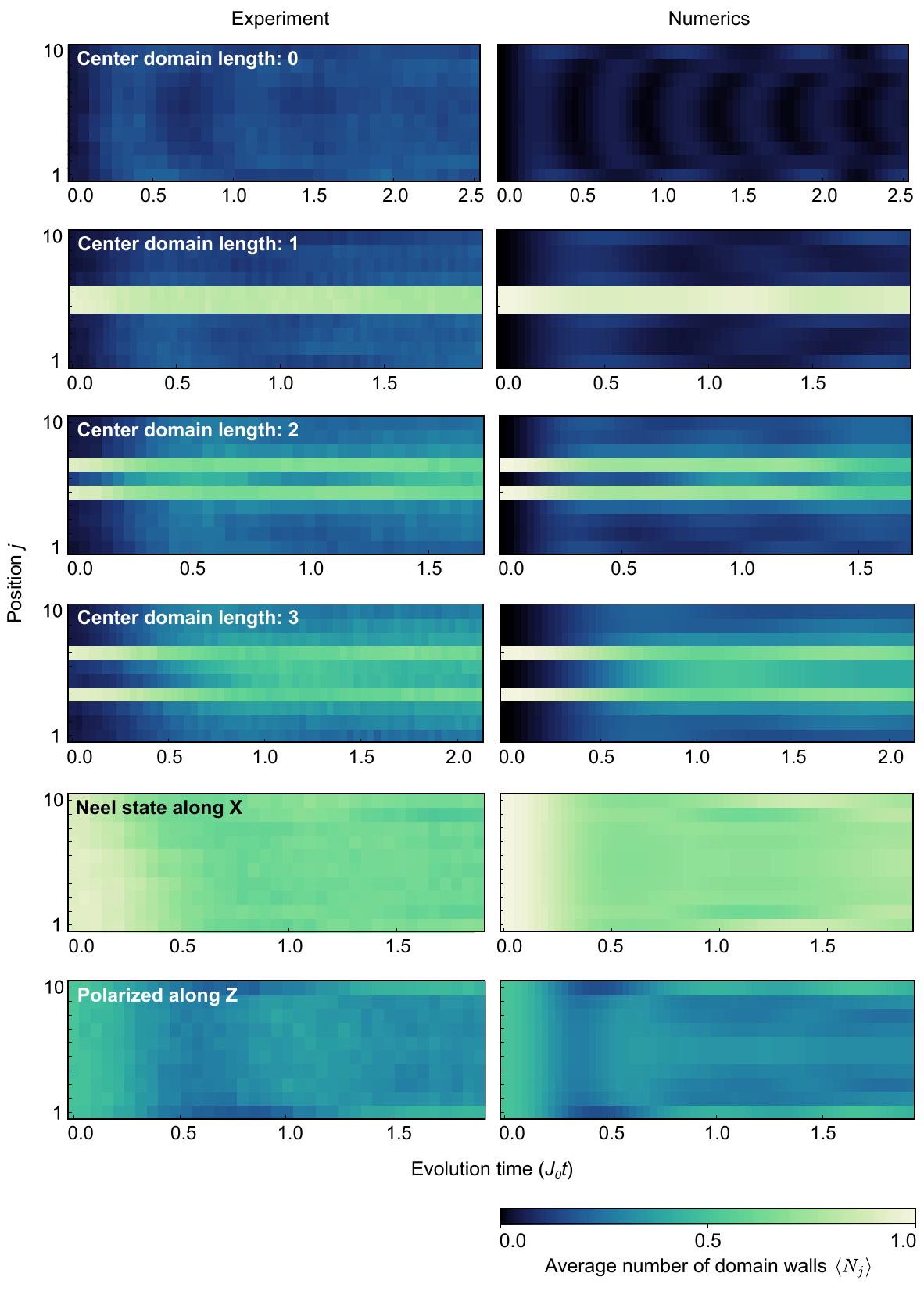}
    \caption{
    \textbf{Domain Wall Localization Under a Confining Hamiltonian Quench.} Evolution of the average number of domain walls $\braket{N_j(t)}$ (\ref{kinkevolution}) for six $L=11$ initial states, each following a quench of the confining Hamiltonian (\ref{ConfinementH}) with $B/J_0 \approx 0.75$. The left column shows experimental data averaged over 2000 experiments and the right column shows numerics calculated by solving the Schr\"{o}dinger equation. Domain wall pairs are prepared by flipping the initial polarization of a central domain of spins. The N\'{e}el state is prepared by flipping the initial magnetization of spins at even-numbered positions.
   }
    \label{fig:KinkLocalization}
\end{figure*}

\renewcommand{\thefigure}{S2}
\begin{figure}[th!]

    \includegraphics[scale=1.0]{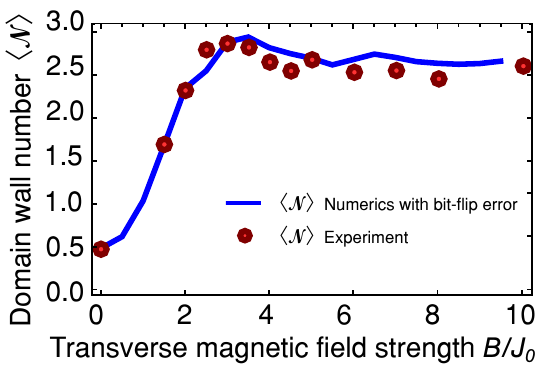}
    \caption{\textbf{Bit-Flip Error Numerical Study in $L=11$ chain for Dynamical Regimes Investigation}. Red dots show the $L=11$ data displayed in Fig. \ref{Fig4}A. The blue line illustrates the numerical value of $\langle\textit{$\mathcal{N}$}\rangle$ with increasing $B$-field, taking bit-flip error into account. We found that a bit-flip error per ion of $2.47\%$ in the numerical calculation matches the experimental data well. The most notable effect of bit-flip errors is an increase in the number of domain walls at $B/J_0=0$ (see Fig. \ref{Fig4}A for comparison with zero bit-flip error numerics).
    }
    \label{fig:BitFlip}
\end{figure}
\renewcommand{\thefigure}{S3}
\begin{figure}[th!]
    \includegraphics[scale=1.0]{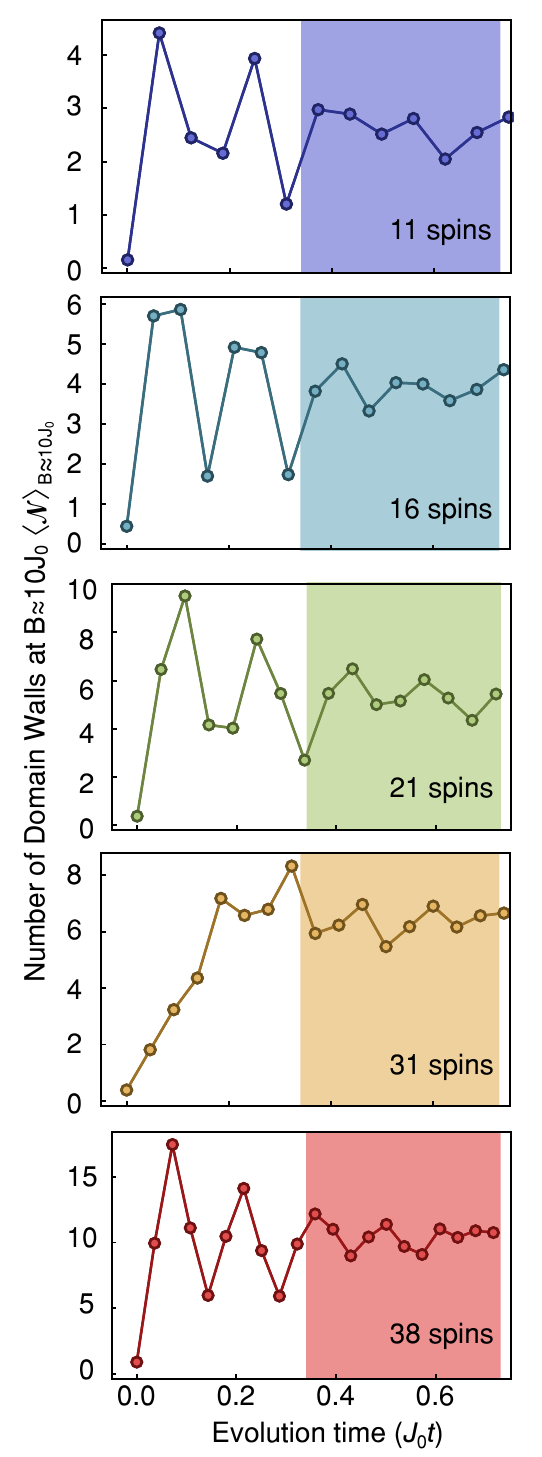}
    \caption{\textbf{Evolution of Domain Wall Population}. Experimental data of evolution of the number of domain walls $\braket{\mathcal{N}}$ during a quench of Hamiltonian (\ref{ConfinementH}) with $B/J_0\approx10$ for multiple system sizes. The shaded area indicates when $\braket{\mathcal{N}}$ converges to a steady state and before qubit dephasing occurs. We fix the scaled integration time $J_0(t_2-t_1)$, although each experiment has a different $J_0$.
    }
    \label{fig:DomainWall}
\end{figure}

\end{document}